\documentclass[10pt]{article} 
\usepackage{amsmath}
\usepackage{cite}
\usepackage{booktabs}
\usepackage{array}
\usepackage{graphicx}
\usepackage{hyperref}
\usepackage{xcolor,colortbl}
\usepackage{pdflscape}

\usepackage{listings}
\usepackage{tikz}
\usepackage{tikz-qtree}
\usetikzlibrary{positioning,shapes,arrows,intersections}
\usepackage{tkz-euclide}
\usetkzobj{all}
\tikzstyle{block} = [rectangle, draw, fill=blue!20, text width=7em, text centered, rounded corners, minimum height=3em]
\tikzstyle{blocka} = [rectangle, draw, fill=blue!20, text width=7em, text centered, minimum height=3em]
\tikzstyle{blockv} = [rectangle, draw, fill=green!20, text width=7em, text centered, rounded corners, minimum height=3em]
\tikzstyle{blockr} = [rectangle, draw, fill=red!20, text width=7em, text centered, rounded corners, minimum height=3em]
\tikzstyle{blockg} = [rectangle, draw, text width=7em, text centered, rounded corners, minimum height=3em]

\tikzstyle{hl} = [draw, rectangle, minimum height=2em, minimum width=3em]
\tikzstyle{subreq}= [draw, rectangle, minimum height=5em, minimum width=11em]
\tikzstyle{proof1}= [draw, rectangle, fill=gray!20,minimum height=2em,minimum width=9em]
\tikzstyle{proof2}= [draw, rectangle, fill=gray!60,minimum height=2em,minimum width=9em]

\usepackage{url}
\usepackage{underscore}

\begin{document}

\title{Verification of Control Systems Implemented in Simulink with Assertion Checks and Theorem Proving: A Case Study}

\author{Dejanira Araiza-Illan\footnote{Department of Computer Science, University of Bristol, Bristol, UK, \newline \href{mailto:dejanira.araizaillan@bristol.ac.uk}{dejanira.araizaillan@bristol.ac.uk}}, Kerstin Eder\footnote{Department of Computer Science, University of Bristol, Bristol, UK, \newline \href{mailto:kerstin.eder@bristol.ac.uk}{kerstin.eder@bristol.ac.uk}}, and Arthur Richards\footnote{Department of Aerospace Engineering, University of Bristol, Bristol, UK, \newline \href{mailto:arthur.richards@bristol.ac.uk}{arthur.richards@bristol.ac.uk}}}

\date{}

\maketitle

\begin{abstract}
This paper presents the verification of control systems implemented in Simulink. The goal is to ensure that high-level requirements on control performance, like stability, are satisfied by the Simulink diagram. A two stage process is proposed. First, the high-level requirements are decomposed into specific parametrized sub-requirements and implemented as assertions in Simulink.  Second, the verification takes place. On one hand, the sub-requirements are verified through assertion checks in simulation. On the other hand, according to their scope, some of the sub-requirements are verified through assertion checks in simulation, and others via automatic theorem proving over an ideal mathematical model of the diagram. We compare performing only assertion checks against the use of theorem proving, to highlight the advantages of the latter. Theorem proving performs verification by computing a mathematical proof symbolically, \emph{covering the entire state space of the variables}. An automatic translation tool from Simulink to the language of the theorem proving tool Why3 is also presented. The paper demonstrates our approach by verifying the stability of a simple discrete linear system.
\end{abstract}

\section{Introduction}
The implementation of control systems before deployment needs to be verified, particularly in domains that involve safety-critical systems~\cite{Meenakshi}. The design of a control system is typically followed by numerical simulation before its final realization as code.  Increasingly, automatic code generation is employed to generate real-time deployment code from the simulation~\cite{DGGM06}, reducing the risk of errors in hand-coding.  In this paper, we propose a method for verifying that the simulation, embodied in a Simulink diagram, satisfies the requirements of the original system.  This method seeks to identify and eliminate errors introduced during the control design in the simulation stage.

Testing and formal methods, namely model checking and theorem proving, have been used to verify systems at different design or implementation levels. None of them are adequate to cover all stages at once in practice. A particular challenge in verifying a control system is the presence of signals and parameters in the domain of the real numbers, which leads to an infinite number of states and state-space explosion problems for some of the verification methods. In testing, samples of the state space of the variables and parameters are used when executing a model of the system, helped by automated search methods as in~\cite{Zhan,Wilmes,Bouissou} for Simulink diagrams.  Model checking is the exhaustive traversal of a model of a system (i.e., enumerating all the states in the model) to check for properties~\cite{ClarkeMC}. Model checking requires a discrete or hybrid state model of the system; it can be affected by the state-space explosion problem. Theorem proving, also denominated deductive verification~\cite{ClarkeMC}, comprises a family of tools to find a mathematical proof of a requirement, automatically via Satisfiability Modulo Theory (SMT) solvers or Satisfiability (SAT) solvers, or interactively (i.e., with additional user guidance). Theorem proving allows generalizing over the entire state-space, as verification problems are posed symbolically. Different kinds of requirements of Simulink diagrams have been verified through model checking~\cite{Chutinan,Meenakshi,Zuliani,Barnat,Proverwebpage} and theorem proving~\cite{Clawz,Circus,Simcheck,Zou,Araiza}. The stability of systems and controllers described mathematically has been verified in theorem provers~\cite{Metitarski,Michel}, but the translation from a mathematical form into the input languages is performed manually. Other approaches verify stability of controllers implemented as C code~\cite{PVS,Zou}, instead of at a design level. Many of the theorem proving approaches for Simulink make use of interactive tools~\cite{Clawz,Circus,Zou,PVS}, whereas fewer use the automation provided by SMT or SAT solvers~\cite{Simcheck,Araiza}. 

In this paper, we present the verification of Simulink diagrams of control systems with respect to high-level design requirements like stability. Our approach involves two stages. First, the high-level requirements are decomposed according to control theory into specific parametrized sub-requirements and implemented as assertions in Simulink. Second, the property is verified. We compare two verification methods to highlight the advantages of theorem proving over simple assertion checks. First, we perform simple assertion checks in simulation. Secondly, we divide the sub-requirements into assertion checks, if they correspond to numerical computations of the constants, and into proof goals to be translated for verification via automatic theorem proving with SMT solvers, if they include system variables. This latter separation into assertion checks and theorem proving was chosen due to the practicality of performing numerical computations in MATLAB instead of SMT solvers, otherwise our intention is to use theorem proving. A large number of assertion checks in simulation would verify a great part of the system's state space against the requirements, but this coverage is not complete. Furthermore, a large number of checks would consume time and memory. A mathematical proof obtained from the verification via theorem proving is symbolic (i.e., for all the possible values of any variable), and an answer can be computed within a reasonable amount of time and memory resources if using SMT solvers. 

Our paper proposes automated translation of Simulink\footnote{Available from: \url{https://github.com/riveras/simulink}} into a language for theorem proving, to reduce errors introduced by manual translation. In particular, we translate the Simulink model into Why, the logic language of the Why3 tool~\cite{Why3man}. This tool translates Why code into the input language of SMT solvers and allows interfacing with them directly, to compute mathematical proofs automatically. In the translation, the Simulink diagram's blocks are captured as relations of their input and output signals. Assertions become verification goals.  We provide Why libraries containing mathematical definitions for some blocks and their properties, to facilitate the translation process. 

This paper demonstrates our verification approach by means of a case study, verifying stability of a discrete linear feedback system.  Our initial results pave the way for more ambitious challenges related to the verification of autonomous systems, such as a model predictive controller (MPC).

\section{Case Study Description and Requirement}\label{System}

Consider a linear discrete system in state-space equation form
\begin{equation}
\mathbf{x}(k+1)=\mathbf{A}\mathbf{x}(k)+\mathbf{Bu}(k) \label{system:loop},
\end{equation}
with the specific values 
\begin{eqnarray}
\mathbf{A}&=&\left[ \begin{array}{cc}
-1.50&-0.50\\ -0.50&-1.00 \end{array}\right], \ \ \mathbf{B}=\left[ \begin{array}{r}
2.00\\ -1.00 \end{array} \right] .
\end{eqnarray}
This system is unstable. We need to design a feedback controller for the system in~(\ref{system:loop}) to satisfy the requirement: \textbf{the closed-loop system shall be stable}. A controller is proposed,
\begin{equation}\label{controller}
\mathbf{u}(k)=-\mathbf{K} \mathbf{x}(k), \ \  \mathbf{K}=\left[\begin{array}{cc} -2.01&-1.32\end{array}\right],
\end{equation}
which has been found by pole placement~\cite{Franklinbook}, with desired poles of $[0.8,-0.6]$.  The implementation of this system and controller in Simulink is shown in the top section of Fig.~\ref{systemreqs}. The constants defining the system and controller are explicitly added as {\footnotesize \texttt{Constant}} blocks, to be treated symbolically in mathematical proofs. 

\section{Verification Process}\label{sec:process}
Lyapunov's second method has been applied directly to determine the stability of the implemented system.  A candidate Lyapunov function is the following:
\begin{equation}
V(\mathbf{x}(k))=\mathbf{x}(k)^{\mathrm{T}}\mathbf{Px}(k),
\end{equation}
where
\begin{equation}
\mathbf{P}= \left[\begin{array}{cc} 22.01&16.93\\ 16.93&15.16\end{array}\right],
\end{equation}
calculated by solving the discrete Lyapunov equation 
\begin{equation}\label{disclyap}
\mathbf{(A-BK)}^{\mathrm{T}}\mathbf{P(A-BK)}-\mathbf{P}=-\mathbf{I}.
\end{equation}

With this information, the requirement can be decomposed as illustrated in Table~\ref{tab:decomp}.  The high-level stability requirement (\textbf{R1}) is decomposed into two child requirements, showing that $V(\mathbf{x}(k))$~is a Lyapunov function for the closed-loop system. Sub-requirement \textbf{R1.1} corresponds to the property of positivity of a Lyapunov function. This sub-requirement can be proved based only on the positive definiteness of the constant matrix~$\mathbf{P}$ (\textbf{R1.1.1}). The requirement for decreasing~$V(\mathbf{x}(k))$ (\textbf{R1.2}) is decomposed into two child requirements: \textbf{R1.2.2}, another property of positive definiteness based on constants; and \textbf{R1.2.1} an identity check of the Lyapunov discrete equation and the Lyapunov's function difference (i.e., the right Lyapunov equation was used to compute matrix $\mathbf{P}$). The latter is more suited to be verified through mathematical proof via theorem proving as, besides avoiding floating point issues, it refers to a comparison of the functionality of sequences of blocks in the Simulink diagram, and it involves system variables with an ample range of values. The SMT solvers, via Why3, would analyze the mathematical description of the Simulink diagram's structure, to find a proof of the identity or not. Verifying all sub-requirements \textbf{R1.1.1}, \textbf{R1.2.1} and \textbf{R1.2.2} would verify the stability requirement. We verify the positive definiteness sub-requirements numerically in Simulink by requiring the minimum eigenvalue to be positive.

\begin{table}[t]
\centering
\footnotesize
\caption{Requirements Decomposition}
\begin{tabular}{|lll|p{6cm}|p{3cm}|}
\hline
\multicolumn{3}{|l|}{\textbf{ID}} & \bf REQUIREMENT & \bf VERIFICATION METHOD \\ \hline
\hline
\multicolumn{3}{|l|}{R1} & The closed-loop system shall be stable. & Decomposed: R1.1 and R1.2 \\ \hline

&\multicolumn{2}{l|}{R1.1} & The function $V(\mathbf{x}(k))$ shall be positive so long as $\mathbf{x}(k) \neq \mathbf{0}.$& Decomposed R1.1.1 \\ \hline

&&R1.1.1&The matrix $\mathbf{P}$ shall be positive definite. & Simulink check \\ \hline

&\multicolumn{2}{l|}{R1.2} & The function $V(\mathbf{x}(k))$ shall be strictly decreasing when $\mathbf{x}(k) \neq \mathbf{0}$.& Decomposed: R1.2.1 and R1.2.2 \\ \hline

&&R1.2.1 & The decrease of the function $V(\mathbf{x}(k))$ shall be equal to $\mathbf{x}(k)^{\mathrm{T}}[\mathbf{P}-\mathbf{(A-BK)}^{\mathrm{T}}\mathbf{P(A-BK)}]\mathbf{x}(k)$. &Theorem prover\\ \hline

&&R1.2.2 & The matrix $\mathbf{P-(A-BK)}^{\mathrm{T}}\mathbf{P(A-BK)}$ shall be positive definite. & Simulink check \\ \hline
\end{tabular}
\label{tab:decomp}
\end{table}

The requirements in the lowest level of the decomposition (e.g.,~\textbf{R1.1.1}, \textbf{R1.2.1} and~\textbf{R1.2.2}) can now be added to the Simulink diagram, attached to {\footnotesize \texttt{Numerical}} or {\footnotesize \texttt{Goal}} blocks, which have {\footnotesize \texttt{Assert}} blocks inside.  Upper level requirements (e.g., \textbf{R1.1} and~\textbf{R1.2}) can be incorporated into the Simulink diagram to add confidence, even though they are verified by their sub-requirements. The requirements that are connected to {\footnotesize \texttt{Numerical}} blocks are verified only through assertion checks in simulation. They should only depend on constants, such that they are proven to always hold. This constraint can be enforced through adding static analysis of the Simulink diagram. {\footnotesize \texttt{Goal}} blocks also contain {\footnotesize \texttt{Assert}} blocks and can be checked in simulation. {\footnotesize \texttt{Goal}} blocks are used as indicators for the translation, being formatted to be verified, whereas {\footnotesize \texttt{Numerical}} blocks are ignored. The remaining blocks in a Simulink diagram are also translated.

\begin{figure}[t]
\centering
\includegraphics[width=\textwidth]{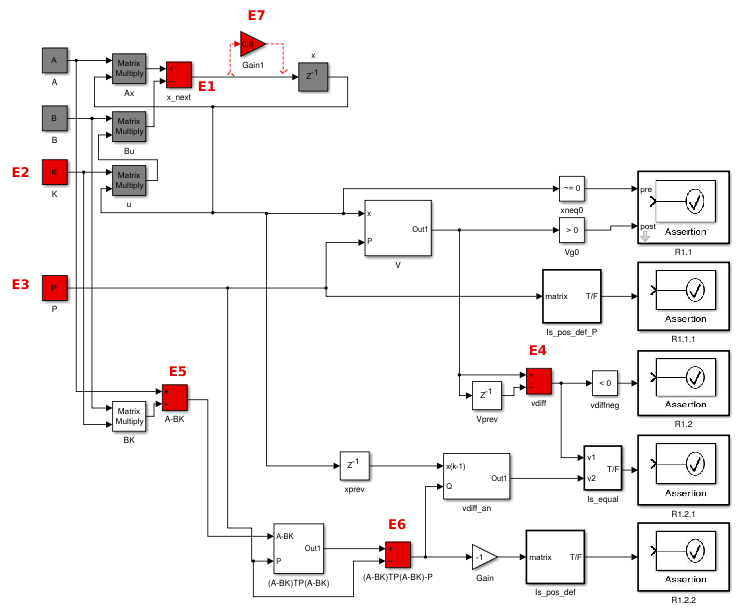} 
\caption{Added Requirements into the Simulink Diagram. System Blocks in Gray. E1--E7: Locations of Introduced Errors.}
\label{systemreqs}
\end{figure} 

The use of numerical methods to compute the eigenvalues for positive definiteness, and any other computation in simulation, are subject to errors due to the use of floating point. For this reason, the identity comparison was encoded as an approximation: $|v_1-V_2|<\epsilon$, where $\epsilon$ is an error bound (e.g., $1\times10^{-5}$). The implications of floating point computations with respect to the satisfaction of the high-level requirements can be investigated in a further analysis level, as in~\cite{Anta} for fixed-point of code implementations of Simulink diagrams.

The number of basic blocks added to the original implementation for the verification is quite significant, as shown in Fig.~\ref{systemreqs}. This trade-off of adding components to perform the verification is well known in the hardware domain, where test-bench code can amount to 80\% of the code in a simulation~\cite{Bergeron}.

\section{Translation Process}\label{translation}
The Simulink diagram is translated automatically to Why, the language of Why3~\cite{Why3man}, to symbolically verify sub-requirements with theorem proving. Why3 automatically translates Why code into the input language of SMT solvers, and controls them directly. The automation of the translation reduces the amount of manual input, and consequently the possibility of introducing errors. The translation procedure has been implemented as a script in MATLAB. The script converts a diagram into an abstract syntax tree~\cite{Simcheck,Araiza} and other data structures, and then follows them when writing the Why code. The translation preserves the semantics. 

In the translation, a logical description of the system's mathematically ideal functionality and properties is represented as a {\footnotesize \texttt{theory}}. This Why file is provided to Why3, and the requested SMT solvers will try to compute a proof of the satisfiability of the verification goals with respect to this system's description. A {\footnotesize \texttt{theory}} is formed by the \textit{signals} connecting the blocks, a logic \textit{description} of the functionality of each block, and the proof \textit{goals} (assertions from the {\footnotesize \texttt{Goal}} blocks).  

Input and output block signals are represented as discrete time functions, mapping an integer to a matrix, in a curried convention. In this paper, we translate all the signals as matrices, an abstraction of vectors and scalars. The signals are named after the block they originated from:
\begin{center}
\begin{minipage}{3.2in} 
\begin{lstlisting}[belowskip=-0.9 \baselineskip]
  function b_block_name int: matrix
\end{lstlisting}
\end{minipage}
\end{center}

The functionality of each block is added by ``cloning'' (or statically linking) a predefined block's mathematical definition (also in a {\footnotesize \texttt{theory}} form), parametrized by the input and output signal names. The definition in Why for a particular block is chosen according to the {\footnotesize\texttt{Block Type}} or {\footnotesize\texttt{MaskType}}. Why definitions (theories) for a set of basic blocks have been developed and are used internally by the translator. The definitions describe the relation between a block's inputs and outputs. For example, the {\footnotesize \texttt{Sum}} block was defined as: 
\begin{center}
\begin{minipage}{3.2in}
\begin{lstlisting} [belowskip=-0.9 \baselineskip]
theory Sum_Add
  use import matrix.Matrix
  function in1 int: matrix
  function in2 int: matrix
  function out1 int: matrix
  axiom def: forall k: int. out1 k = m_sum (in1 k) (in2 k)
end
\end{lstlisting}
\end{minipage}
\end{center}

The produced {\footnotesize \texttt{theory}} in Why for the system in this paper has the structure:
\begin{center}
\begin{minipage}{3.2in}
\begin{lstlisting} [belowskip=-0.9 \baselineskip]
theory System_controller
  use import matrix.Matrix
  function b_a int: matrix
  ...
  clone Sum_Add as B_x_next with function in1 = b_ax, function in2 = b_bu, function out1 = b_x_next
  ...
  goal G_b_compare: forall k:int. v_diff k = vdiff_an k 
  ...
end
\end{lstlisting}
\end{minipage}
\end{center}

We dealt with the presence of subsystems by translating them as individual theories. Otherwise, they can be masked and considered as basic blocks, with their respective definitions added to the library for the translator.

Additionally, we have developed a linear algebra {\footnotesize{\texttt{theory}}} in Why, containing our definitions of matrix, algebraic operations and properties (e.g., commutativity, associativity), based on other linear algebra theories~\cite{Nakamura,PVS,Shi}. Our definitions are based on real numbers, corresponding to an ideal mathematical model of the system in the Simulink diagram. For example, matrix multiplication, and an associativity of matrix multiplication axioms are defined as:
\begin{center}
\begin{minipage}{3.2in}
\begin{lstlisting}[belowskip=-0.9 \baselineskip]
  function mxm matrix matrix : matrix
  axiom assoc_mult: forall m1 m2 m3:matrix. 
  		mxm (mxm m1 m2) m3 = mxm m1 (mxm m2 m3)
\end{lstlisting}
\end{minipage}
\end{center}

\section{Evaluation and Results} \label{results}

\begin{figure}[t]
\centering
\includegraphics[scale=0.5]{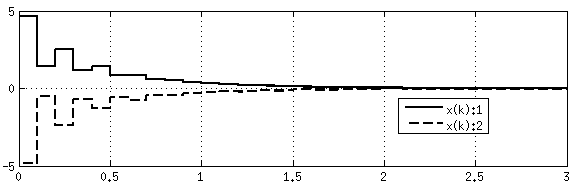} 
\caption{Controlled System's State Variables $\mathbf{x}(k)$.}
\label{systemsignals}
\end{figure} 

The correct Simulink diagram of the designed control system yields a stable behaviour and convergence of the state to an equilibrium point in the origin, as shown in Fig.~\ref{systemsignals}. We compared performing verification with simple assertion checks in simulation at different sub-requirement decomposition levels, against verifying based on a separation of assertions about constant parameters and assertions about variables tackled through symbolic mathematical proof via theorem proving. This comparison allows commenting on the importance of a more comprehensive methodology for the verification of a Simulink diagram, and showcases the advantages of our use of theorem proving to automatically verify some of the sub-requirements, for the entire state-space of the variables.

Errors have been injected into the Simulink's components of the system or verification, as identified in Fig.~\ref{systemreqs}:

\begin{enumerate}
\renewcommand{\labelenumi}{E\arabic{enumi}.}
\item The feedback is reversed by using an {\footnotesize\texttt{Add}} block instead of {\footnotesize\texttt{Subtract}} for {\footnotesize\texttt{x_next}}, leading to $\mathbf{A+BK}$ and making the system unstable. $\mathbf{P}$ and $\mathbf{K}$ are computed for $\mathbf{A-BK}$.
\item $\mathbf{K}$ is computed for $\mathbf{A+BK}$ but we implement $A-BK$. This error can happen when loading $\mathbf{-K}$ into the workspace by mistake. 
\item Using a positive definite matrix $\mathbf{P}$ computed with $(\mathbf{A-BK})^{\mathrm{T}}$ and not $\mathbf{A-BK}$.  This error is likely to happen as MATLAB's {\footnotesize\texttt{dlyap}} function requires the transposition of the system's parameters. 
\item Inversion of the difference $V(\mathbf{x}(k))-V(\mathbf{x}(k-1))$ to $V(\mathbf{x}(k-1))-V(\mathbf{x}(k))$. This is achieved by swapping the inputs to the corresponding {\footnotesize\texttt{Subtract}} block.
\item Swapping a {\footnotesize\texttt{Subtract}} block for an {\footnotesize\texttt{Add}} block when computing the predicted Lyapunov function change. This means assuming positive feedback instead of negative feedback. 
\item Inverting the sign in the predicted difference of the Lyapunov function by swapping the inputs of the corresponding {\footnotesize\texttt{Subtract}} block, resulting in $\mathbf{P-(A-BK)}^{\mathrm{T}}\mathbf{P(A-BK)}$ instead of $\mathbf{(A-BK)}^{\mathrm{T}}\mathbf{P(A-BK)-P}$.
\item Extra {\footnotesize\texttt{Gain}} block in the system, such that $\mathbf{x}(k+1)=0.8\mathbf{(A-BK)x}(k)$ is a stable system but different to the one used for design. 
\end{enumerate}

The first two errors are used to illustrate the results of the verification approach when the system is unstable and a Lyapunov function is proposed. The other five errors show that errors in the added assertions do not lead to false conclusions about the high-level requirement being verified, stability.  

We used an Intel i5-3230M 2.60\,GHz CPU, with 8\,GB of RAM, running Ubuntu 14.04, and Simulink in MATLAB R2013a.  We used the Z3 (version 4.3.1)~\cite{z3url} and CVC4 (version 1.4)~\cite{cvc4url} SMT solvers controlled through Why3 (version 0.81). Why files were produced automatically using the translation of Section~\ref{translation}.  

\begin{landscape}
\begin{table}[t]
\centering
\caption{Evaluation Results}
\label{tab:evalresults}
\begin{tabular}{|l|l|c|c|c|c|c|c|c|c|}
\hline
\bf REQ. & \bf METHOD & \bf CORR.& \bf E1 & \bf E2 & \bf E3 & \bf E4 & \bf E5 &\bf E6 & \bf E7\\
\hline
\multicolumn{10}{|c|}{\textbf{Checking Assertions for Different Initial Conditions in Upper and Lower Sub-requirement Levels}}\\
\hline
R1.1 & Simulink check: $\mathbf{x}(0)=[1,1]^{\mathrm{T}}$ 		& Pass & Pass & Pass & Pass & Pass & Pass & Pass & Pass\\
R1.1 & Simulink check: $\mathbf{x}(0)=[1,-0.8]^{\mathrm{T}}$	& Pass & Pass & Pass & Pass & Pass & Pass & Pass & Pass\\
\hline
R1.2 & Simulink check: $\mathbf{x}(0)=[1,1]^{\mathrm{T}}$ 		& Pass & FAIL & FAIL & FAIL & FAIL & Pass & Pass & Pass \\
R1.2 & Simulink check: $\mathbf{x}(0)=[1,-0.8]^{\mathrm{T}}$ 	& Pass & FAIL & FAIL & Pass & FAIL & Pass & Pass & Pass \\
\hline
R1.1.1 & Simulink check (any initial condition) & Pass & Pass & Pass & Pass & Pass & Pass & Pass & Pass\\
\hline
R1.2.1 & Simulink check: $\mathbf{x}(0)=[1,1]^{\mathrm{T}}$ 	& Pass & FAIL & FAIL & Pass & FAIL & FAIL & FAIL & FAIL \\
R1.2.1 & Simulink check: $\mathbf{x}(0)=[1,-0.8]^{\mathrm{T}}$& Pass & FAIL & FAIL & Pass & FAIL & FAIL & FAIL & FAIL \\
\hline
R1.2.2 & Simulink check (any initial condition) 	& Pass & Pass & FAIL & FAIL & Pass & FAIL & FAIL & Pass\\
\hline
\multicolumn{10}{|c|}{\textbf{Division of Sub-requirements in Lowest Level into Assertion Checks and Theorem Proving}}\\
\hline
R1.1.1 & Simulink check										& Pass & Pass & Pass & Pass & Pass & Pass & Pass & Pass \\
\hline
R1.2.1 & Theorem proving 									& Pass & FAIL & Pass & Pass & FAIL & FAIL & FAIL & FAIL \\
\hline
R1.2.2 & Simulink check 									 	& Pass & Pass & FAIL & FAIL & Pass & FAIL & FAIL & Pass \\
\hline
\end{tabular}
\end{table}
\end{landscape}

Table~\ref{tab:evalresults} shows the verification results. The column labelled {\footnotesize\texttt{CORR.}} refers to the nominal stable system and a correctly applied Lyapunov's second method. Columns {\footnotesize\texttt{E1-E7}} show the verification results for systems with errors. In simulation, a {\footnotesize\texttt{Pass}} means the assertion passed the check, whereas {\footnotesize\texttt{FAIL}} indicates the assertion failed the check. In theorem proving, {\footnotesize\texttt{Pass}} indicates that both SMT solvers found the assertion valid, which is expected after less than 10 seconds. {\footnotesize\texttt{FAIL}} indicates the failure to prove the assertion: the SMT solvers timed out after 100 seconds, or ran out of memory within 100 seconds. 

For the verification based on simple assertion checks, two initial conditions, $\mathbf{x}(0)$, were applied and simulated for 10 seconds. First, the assertions corresponding to the first decomposition level in Table~\ref{tab:decomp}, \textbf{R1.1} and \textbf{R1.2}, were checked. Then, sub-requirements \textbf{R1.1.1}, \textbf{R1.2.1} and \textbf{R1.2.2} were checked, to highlight the advantages of the decomposition. False positives of the high-level stability requirement are likely to emerge for errors E3 and E5--E7, if only (\textbf{R1.1} and \textbf{R1.2}) are checked: the Lyapunov function is positive and decreases for the chosen initial conditions, leading the designers to conclude the system is stable. This lack of information about the system can be corrected through simulations for more initial conditions. Alternatively, the lowest level of sub-requirements can be considered instead. Sub-requirements \textbf{R1.1.1} and \textbf{R1.2.2} provide information about the system's constant parameters that does not depend on the system's variables. Sub-requirement \textbf{R1.2.1} provides information about the coherence of the system's structure with respect to the Lyapunov's second method. The same false positives are avoided if checking sub-requirements \textbf{R1.1.1}, \textbf{R1.2.1} and \textbf{R1.2.2} instead, for the same initial conditions. Nevertheless, checking requirement \textbf{R1.2.1} numerically in simulation is subject to floating point issues. These numerical issues, present in the verification of sub-requirements about variables in the system, can be alleviated by the use of theorem proving, due to its symbolic nature. 

For the second verification alternative, the verification of sub-requirements \textbf{R1.1.1}, \textbf{R1.2.1} and \textbf{R1.2.2} is split into assertion checks and mathematical proof via theorem proving, as means to verify the high-level requirement if all the assertions are verified as valid. For requirements \textbf{R1.1.1} and \textbf{R1.2.2}, we considered the results of the previous assertion checks, since they do not depend on initial conditions nor simulation time. The results in Table~\ref{tab:evalresults} indicate that every error is detected, as the verification of at least one the three sub-requirements fails. All the error implementations preserved the positive definiteness of the matrix~$\mathbf{P}$ in the candidate Lyapunov function, so requirement \textbf{R1.1.1} is satisfied. The analysis of requirement \textbf{R1.2.1} by theorem proving fails for errors E1 and E4--E7, where the assertions are not consistent with the system. The implementations with errors E2 and E3 are valid according to theorem proving, as sub-requirement (\textbf{R1.2.1}) is verified symbolically (i.e., without considering the values of the matrices in the system). Errors E2 and E3 refer to matrices $\mathbf{P}$~and~$\mathbf{K}$, and the numerical analysis detects them instead. Errors E3--E6 demonstrate the importance of adding assertions correctly. 

The use of formal methods, in particular theorem proving, allows the verification of sub-requirements for all possible state-space values and parameter combinations at once, due to their inherent symbolic nature. This is advantageous compared to running a large amount of simulations, which can never be exhaustive for systems with variables in the domain of the real (or floating point) numbers. Formal methods can be complemented with simulation-based testing (i.e., running a comprehensive set of assertion checks in simulation), as it can be used to find examples in the state-space that cause the failure of the assertions. Also, simulation-based testing can help to establish if an implemented control system fulfils its purpose with a target environment. 

Since the sub-requirements added as assertions in Simulink make use of default blocks, other tools for testing and model checking such as Simulink Design Verifier~\cite{SDV} can complement the theorem proving verification approach. Also, the assertions can be exported as part of automatically generated code, to act as monitors at runtime. 

\section{Conclusions}\label{conclusion}

In this paper, we presented an approach to verify high-level properties of control systems represented as Simulink diagrams. We use popular implementation and verification tools like Simulink and Why3. The approach provides more automation in the verification process, through the automatic translation of the Simulink diagrams for theorem proving, and the use of automatic theorem provers. The approach can be extended to verifying a variety of high-level properties for different systems, and can be complemented by other verification methods, such as simulation-based testing. The verification of errors in the Simulink diagrams presented in this paper illustrates the need for verification of control systems designs, before code generation. 

We have exemplified our approach through the stability analysis of a controlled discrete system, using Lyapunov's second method. We showed how to derive sub-requirements from a high-level requirement such as stability. We also showed how to add these sub-requirements into the Simulink diagram, using Simulink's default blocks. Although the sub-requirements can be verified in simulation, using assertion checks in Simulink, a more efficient and comprehensive option is separating the verification of the sub-requirements into assertion checks (for constant parameters), and automatic theorem proving via SMT solvers (for system variables). The Simulink diagram can be automatically translated into Why (the logic language of the Why3 tool), to reduce errors from manual input. The use of theorem proving combined with numerical checks about the system's constants, provides a stronger set of assurances about the high-level requirements of the system, since theorem proving is symbolic and results in a mathematical proof that covers the entire state space of variables. Comparatively, assertion checks are not exhaustive over variables or parameter choices, but they can be used to complement theorem proving by providing examples of the violation of a requirement. 

We want to apply the method and tools to more complex systems in the near future, such as an MPC with automatically derived Lyapunov functions. Also, we are expanding the amount of supported Simulink blocks to be used in diagrams, and our linear algebra library.

\section*{Acknowledgements}
The work presented in this paper was supported by the EPSRC grant EP/J01205X/1 RIVERAS: Robust Integrated Verification of Autonomous Systems. The authors thank the reviewers for their comments.

\bibliographystyle{plain}

\bibliography{references}

\begin{thebibliography}{10}

\bibitem{Anta}
Adolfo Anta, Rupak Majumdar, Indranil Saha, and Paulo Tabuada.
\newblock Automatic verification of control systems implementations.
\newblock In {\em Proc. EMSOFT}, pages 9--18, Scottsdale, Arizona, USA, 2010.
  ACM.

\bibitem{Araiza}
D.~{Araiza-Illan}, K.~Eder, and A.~Richards.
\newblock Formal verification of control systems' properties with theorem
  proving.
\newblock In {\em Proceedings of UKACC CONTROL}, pages 244--249, Loughborough,
  UK, 2014. IEEE.

\bibitem{Clawz}
R.~Arthan, P.~Caseley, C.~{O'Halloran}, and A.~Smith.
\newblock {ClawZ}: control laws in {Z}.
\newblock In {\em Proc. ICFEM}, pages 169--176, York, UK, 2000.

\bibitem{Barnat}
J.~Barnat, J.~Beran, L.~Brim, T.~Kratochv\'{i}la, and P.~Ro\u{c}kai.
\newblock Tool chain to support automated formal verification of avionics
  {Simulink} designs.
\newblock In {\em Formal Methods for Industrial Critical Systems}, pages
  78--92, Paris, France, 2012.

\bibitem{Bergeron}
Janick Bergeron.
\newblock {\em Writing Testbenches: Functional Verification of {HDL} Models}.
\newblock Springer, 2003.

\bibitem{Why3man}
F.~Bobot, J.C. Filli\^{a}tre, C.~March\'{e}, G.~Melquiond, and A.~Paskevich.
\newblock {\em The {Why3} Platform}.
\newblock University Paris-Sud, CNRS, Inria, March 2013.

\bibitem{Bouissou}
O.~Bouissou, S.~Mimram, B.~Strazzulla, and A.~Chapoutot.
\newblock Set-based simulation for design and verification of {Simulink}
  models.
\newblock In {\em Proc. ERTS}, Tolouse, France, 2014.

\bibitem{Circus}
A.~Cavalcanti and P.~Clayton.
\newblock Verification of control systems using {Circus}.
\newblock In {\em Proc. ICECCS}, Stanford, CA, USA, 2006.

\bibitem{Chutinan}
A.~Chutinan and B.H. Krogh.
\newblock Computational techniques for hybrid system verification.
\newblock {\em IEEE Transactions on Automatic Control}, 48(1):64--75, January
  2003.

\bibitem{ClarkeMC}
E.~M. Clarke, O.~Grumberg, and D.~A. Peled.
\newblock {\em Model Checking}.
\newblock MIT Press, USA, 1999.

\bibitem{DGGM06}
S.~D'Amico, E.~Gill, M.~Garcia, and O.~Montenbruck.
\newblock {GPS}-based real-time navigation for the {PRISMA} formation flying
  mission.
\newblock In {\em Proc. NAVITEC}, Noordwijk, The Netherlands, December 2006.

\bibitem{Metitarski}
W.~Denman, M.H. Zaki, S.~Tahar, and L.~Rodrigues.
\newblock Towards flight control verification using automated theorem proving.
\newblock In {\em NASA Formal Methods}, pages 89--100, Pasadena, CA, USA, 2011.

\bibitem{Franklinbook}
G.~Franklin, J.D. Powell, and A.~{Emami-Naeini}.
\newblock {\em Feedback Control of Dynamic Systems}.
\newblock Prentice Hall, 2011.

\bibitem{PVS}
H.~{Herencia-Zapana}, R.~Jobredeaux, S.~Owre, P.L. Garoche, E.~Feron, G.~Perez,
  and P.~Ascariz.
\newblock {PVS} linear algebra libraries for verification of control software
  algorithms in {C/ACSL}.
\newblock In {\em NASA Formal Methods}, pages 147--161, Norfolk, VA, USA, 2012.

\bibitem{cvc4url}
http://cvc4.cs.nyu.edu/web/.

\bibitem{z3url}
http://z3.codeplex.com.

\bibitem{Meenakshi}
B.~Meenakshi, A.~Bhatnagar, and S.~Roy.
\newblock Tool for translating {Simulink} models into input language of a model
  checker.
\newblock In {\em Formal Methods and Software Engineering}, pages 606--620,
  Macao, China, 2006.

\bibitem{Michel}
L.~Michael.
\newblock Bernstein-based polynomial approach to study the stability of
  switched systems and formal verification using {HOL Light}.
\newblock {\em arXiv:1410.3969}, 2014.

\bibitem{Nakamura}
Y.~Nakamura, N.~Tamura, and W.~Chang.
\newblock A theory of matrices of real elements.
\newblock {\em Formalized Mathematics}, 14(1):21--28, 2006.

\bibitem{Proverwebpage}
Prover.
\newblock Prover plug-in.

\bibitem{Simcheck}
P.~Roy and N.~Shankar.
\newblock {SimCheck:} a contract type system for {Simulink}.
\newblock {\em Innovations in Systems and Software Engineering}, 7:73--83,
  2011.

\bibitem{Shi}
Z.~Shi, Z.~Liu, Y.~Guan, S.~Ye, J.~Zhang, and H.~Wei.
\newblock Formalization of function matrix theory in {HOL}.
\newblock {\em Journal of Applied Mathematics}, 2014:1--10, 2014.

\bibitem{SDV}
{The Mathworks}.
\newblock {Simulink Design Verifier}, 2012.

\bibitem{Wilmes}
B.~Wilmes.
\newblock Automated structural testing of {Simulink}/{TargetLink} models via
  search-based testing assisted by prior-search static analysis.
\newblock In {\em Proc. VALID}, pages 51--56, Lisbon, Portugal, 2012.

\bibitem{Zhan}
Y.~Zhan and J.A. Clark.
\newblock A search-based framework for automatic testing of {MATLAB}/{Simulink}
  models.
\newblock {\em The Journal of Systems and Software}, 81:262--285, 2008.

\bibitem{Zou}
L.~Zou, N.~Zhan, S.~Wang, M.~Fr\"{a}nzle, and S.~Qin.
\newblock Verifying {Simulink} diagrams via a hybrid {Hoare} logic prover.
\newblock In {\em Proc. EMSOFT}, pages 1--10, 2013.

\bibitem{Zuliani}
P.~Zuliani, A.~Platzer, and E.M. Clarke.
\newblock Bayesian statistical model checking with application to
  {Stateflow}/{Simulink} verification.
\newblock {\em Formal Methods in System Design}, 43(2):338--367, October 2013.

\end{thebibliography}

\end{document}